\documentclass[preprint,12pt]{elsarticle}
\usepackage{amssymb}
\usepackage{epsfig}
\usepackage{amsthm}
\usepackage{amsmath}
\journal{Chemical Physics Letters}
\begin{document}
\begin{frontmatter}

\title{Exploring the relaxation time, bond length and length of the polymer chain in the kinetics of end-to-end looping of a long polymer chain.}

\author{Moumita Ganguly\corref{cor1}\fnref{fn1}}
\ead{mouganguly09@gmail.com}
\address{Indian Institute of Technology Mandi, Mandi (H.P.), India. Pin-175001}

\author{Anirudhha Chakraborty\corref{cor2}}
\address{Indian Institute of Technology Mandi, Mandi (H.P.), India. Pin-175001}
\cortext[cor1]{Corresponding author}

\fntext[fn1]{Phone:-+91-190-5267067, Fax: +91-1905-237945}

\begin{abstract} 
We present a theoretical model for understanding the kinetics of a long  polymer in solution. A Smoluchowski-like equation has been used to model the problem of polymer looping with a Dirac delta function sink of finite strength. The results for rate constants have been obtained by using Green's function. In this model, the rate constants for end-to-end looping of polymer exhibits simple Arrhenius behavior. The rate constants are sensitive to relaxation time, bond length and chain length dependence of the looping probability. In particular, for short term rate constant, relaxation time makes loop formation between two ends of a polymer chain slower but has a direct dependence with long term rate constant. We also show that for shorter polymer strands looping is slower as compared to polymer having longer chain length. The long term rate constant shows inverse square root dependence on the length of the chain. The bond length strongly affects the probability of loop formation as for long constructs of polymer, the end-to-end distance is found to exhibit inverse power law relationship to bond length. We also observe that the short term rate constant for flexible polymer shows stronger affinity for loop formation.
\end{abstract}

\begin{keyword}
Looping, Relaxation time, Bond length, Chain length, Analytical

\end{keyword}

\end{frontmatter}
\section{Introduction}
\noindent The dynamics of looping in polymer has been known for many years and is of great interest in biophysics. It is one of the important reaction step in folding events of biopolymers viz. protein \cite {Buscaqlia} and RNA folding \cite {Thirumalai}. The first part of this letter gives a general procedure for solving the looping dynamics of long polymer molecule in solution by using a Smoluchowski-like equation. Explicite expressions for two different rate constants $k_{I}$ and $k_{L}$ have been obtained. Following this we try to get a coherent understanding of relative roles of relaxation time, length of the polymer chain and bond length on short term and long term rate constants. In our previous paper \cite{Mou}, we solve for the discrete equation for polymer which condenses to motion of a single particle with harmonic oscillator. We make use of one dimensional description of a polymer as given by Szabo {\it et.al.,} \cite{Szabo}. The polymer is composed of a total $2N$ segments of unit length and $x$ is the end-to-end distance, with the value $x=2j$. The polymer exhibits unbiased random walk either in left or right direction and thus can achieve any of $2^{2N}$ different conformations. Thus after N steps, the equilibrium end-to-end distribution $P_{0j}$ of the polymer is given by
\begin{equation}
P_{0j}=2^{-2N}(^{2N}_{N+j})
\end{equation}
where $N+j$ and $N-j$ are the number of polymer segments in right and left direction respectively. Now considering the variation of all left and right monomer segments being independent of each other, the fluctuation of the polymer molecule in solution is given by the following rate equation.
\begin{equation}
\label{2}\frac{d}{dt}\left[^p_n\right]=\frac{1}{\tau_R}\left[
\begin{array}{cc}
-1&1\\
1&1
\end{array}
\right]\left[^p_n\right],
\end{equation}
where the vector $[^p_n]$ accounts for the activity of right and left segments orientations. $\tau_R$ represents the relaxation time which is required to go from one configuration to another. The individual monomer can reorient in $2N$ ways to swap itself from a $x=2j$ conformation either to a $x=2j+2$or $x=2j-2$ conformation and $N-J+1$ ways to reorient a $x=2j+2$ conformation into a $x=2j$ conformation.
Now the resulting master equation for the end-to-end distribution P(j,t)in the (2N+1)-dimensional space is given by \cite{Szabo}
\begin{equation}
\tau_R\frac{d}{dt}P(j,t)=-2NP(j,t)+(N+j+1)P(j+1,t)+(N-j+1)P(j-1,t).
\end{equation}
\noindent For a long chain molecule ($N$ large), the equilibrium distribution of  Eq.(1) can be approximated by a continuous Gaussian distribution ($x = 2 b j$)
\begin{equation}
\label{de}
P_{0}(x)=\frac{e^{-\frac{x^2}{4 b^2 N}}}{(4 \pi b^2 N)^{1/2}}
\end{equation}
Now in continuum limit when each of the monomers are further reduced to close vicinity, then the corresponding probability conservation equation is given by the following equation
\begin{equation}
\tau_{R}\frac{\partial P(x,t)}{\partial t} = \left(4 N b^2 \frac{\partial^2}{\partial x^2} + 2 \frac{\partial}{\partial x} x \right) P(x,t).
\end{equation}
Here the term `$b$' denotes the bond length of the polymer. The term $P(x,t)$ represents for the probability density of end-to-end distance at time $t$. In our model, the effect of various topology classes of polymer loops formed from any end to the interior of the polymer is given by the term $k_{s} P(x,t)$  thus incorporating the effects of all other chemical reactions apart from the end-to-end loop formation. Hence,we get the following modified Smoluchowski equation
\begin{equation}
\tau_{R}\frac{\partial P(x,t)}{\partial t} = \left(4 N b^2 \frac{\partial^2}{\partial x^2} + 2 \frac{\partial}{\partial x} x  - k_{s}\right) P(x,t).
\end{equation}
\noindent Now at $x = 0$, the two ends of the polymer molecule meet and a loop is formed. The $x$ dependent sink term $S(x)$ takes care of the occurrence of the looping reaction (taken to be normalized {\it i.e.} $\int_{-\infty}^{\infty} S(x)dx = 1 $) in the following equation. 
\begin{equation}
\tau_{R}\frac{\partial P(x,t)}{\partial t} = \left(4 N b^2 \frac{\partial^2}{\partial x^2} + 2 \frac{\partial}{\partial x} x - k_{s} - S(x)  \right) P(x,t).
\end{equation}
The rate $k_{s}$ is the rate of loss of probability of end-to-end distance.

\section{Exact result:}
\noindent To obtain the exact result, we do the Laplace transform $\tilde P(x,s)$ of $P(x,t)$ of Eq.(7)which yields\\
\begin{equation}
\left[s  - {\cal L}+ \frac{1}{\tau_{R}} S(x)+\frac{k_{s}}{\tau_{R}}\right] {\tilde P}(x,s)=  P(x,0).
\end{equation}
In which the operator ${\cal L}$ is defined as follows
\begin{equation}
{\cal L} = \frac{4 N b^2}{\tau_{R}} \frac{\partial^2}{\partial x^2} + \frac{2}{\tau_{R}} \frac{\partial}{\partial x} x.
\end{equation}
The solution of this equation in terms of Green's function $G(x,s|x_0)$ can be written as
\begin{equation}
\tilde P(x,s)= \int^\infty_{-\infty} dx_{0}G(x,s+\frac{k_{s}}{\tau_{R}}|x_0)P(x_0,0),
\end{equation}
where $G(x,s|x_0)$ is given by the following equation
\begin{equation}
\left[s - {\cal L} + \frac{1}{\tau_{R}}S(x)\right]G(x,s|x_0)=\delta(x-x_0).
\end{equation}
Using the operator representations of quantum mechanics, we solve and get the following
\begin{equation}
G(x,s|x_0)=\langle x|[x - {\cal L}]^{-1}|x_0\rangle -\langle x|[s - {\cal L}]^{-1}\frac{1}{\tau_{R}}S[s - {\cal L} + \frac{1}{\tau_{R}} S[s - {\cal L} + \frac{1}{\tau_{R}} S]^{-1}|x_0\rangle.
\end{equation}
where $|x\rangle (|x_0\rangle)$ implies the position eigen-ket.
Inserting the intent of identity I= $\int^\infty_{-\infty} dy |y \rangle \langle y|$ in the appropriate places in the second term of the above equation, we get the following equation. 
\begin{equation}
G(x,s|x_0)= G_{0}(x,s|x_0) - \int^{\infty}_{-\infty} dy G_{0}(x,s|y)S(y)G(y,s|x_0)
\end{equation}
\noindent $G_{0}(x,s|x_0)$ is defined by
\begin{equation}
G_{0}(x,s|x_0)=\langle x|[x - {\cal L}]^{-1}|x_0\rangle
\end{equation}
and corresponds to change in end-to-end distance of the polymer, that has the inceptive value $x_0$, in the absence of any sink. It is noteworthy that Laplace transform of $G_0(x,t|x_0)$ gives the probability that the end-to-end distance of a polymer per say, starting at $x_0$ may be found at $x$, at time $t$. It obeys the following equation,
\begin{equation}
\left[(\partial /\partial t)- {\cal L} \right] g_0(x,t|x_0)=\delta(x-x_0).
\end{equation}
The above equation doesn't have a sink term in it. In the absence of sink, there is no absorption of the particle. Therefore, $\int^\infty_{-\infty} dx g_0(x,t|x_0) = 1$. From this we can conclude 
\begin{equation}
\int^\infty_{-\infty} dx G_0(x,s|x_0) = 1/s
\end{equation}
If $S(y)= k_{0}\delta(y - x_s)$, then Eq.(16) becomes \\
\begin{equation}
G(x,s|x_0)=G_0(x,s|x_0) - \frac{k_0}{\tau_{R}}\;G_0(x,s|x_s)G(x_s,s|x_0).
\end{equation}
We now solve Eq.(20) to find
\begin{equation}
G(x_s,\;s|x_0)=G_0(x_s,\;s|x_0) [1+ \frac{k_0}{\tau_{R}} G_0(x_s,\;s|x_s)]^{-1}.
\end{equation}
When substituting the above back into Eq.(20) gives
\begin{equation}
G(x,s|x_0)=G_0(x,s|x_0)- \frac{k_0}{\tau_{R}} G_0(x,s|x_s)G_0(x_s,s|x_0)[1+ \frac{k_0}{\tau_{R}} \;G_0(x_s,s|x_s)]^{-1}.
\end{equation}
Using the expression of $G(x,s|x_0)$ in Eq.(11) we get $\tilde P(x,s)$ explicitly. It is difficult to calculate survival probability $P_e(t) =\int^\infty_{-\infty} dx P(x,t)$. Instead one can easily calculate the Laplace transform $P_e(s)$ of $ P_e(t)$ directly. $P_e(s)$ is associated to $P(x,s)$ by 
\begin{equation}
P_e(s) = \int^\infty_{-\infty} dx {\tilde P}(x,s).
\end{equation} 
From Eq. (11), Eq. (22) and Eq. (23), we get
\begin{equation}
P_e(s)=\frac{1}{s+k_{s}}\left[1-[1+\frac{k_0}{\tau_{R}} G_0(x_s,s+k_{s}|x_s)]^{-1} \frac{k_0}{\tau_{R}} \; \times \int^\infty_{-\infty} dx_0 \; G_0 (x_s,s+k_{s}|x_0)P(x_0,0).\right]
\end{equation}
The average and long time rate constants can be derived from $P_e(s).$ Thus, $k^{-1}_I =P_e(0)$ and $k_L$ = negative of the pole of $P_e(s),$ which is close to the origin. From (23), we obtain
\begin{equation}
k^{-1}_I =\frac{1}{k_{s}}\left[1- [1+\frac{k_0}{\tau_{R}} G_0(x_s,k_{s}|x_s)]^{-1} \frac{k_0}{\tau_{R}} \; \times \int^\infty_{-\infty} dx_0 \; G_0 (x_s,k_{s}|x_0)P(x_0,0).\right]
\end{equation}
Thus $k_I$ depends on the initial probability distribution $P(x,0)$ whereas $k_L = - $ pole of $[\;[ 1+\frac{k_0}{\tau_R}\; G_0(x_s, s+k_s|x_s)][s+k_s]\;]^{-1}$, the one which is closest to the origin, on the negative $s$ - axis, and is independent of the initial distribution $P(x_0,0)$.
The $G_0(x,s;x_0)$ can be found out by using the following equation {\cite{KLS}:
\begin{equation}
\left(s - {\cal L}\right) G_{0}(x,s;x_0)= \delta (x - x_0) 
\end{equation}
Using standard method \cite{Hilbert} to obtain.
\begin{equation}
G_0(x,s;x_0)=F(z,s;z_0)/(s+k_s)
\end{equation}
with
\begin{equation}
F(z,s;z_0)= D_\nu(-z_<)D_\nu(z_>)e^{(z_0^2-z^2)/4}\Gamma(1-\nu)[1/(4 \pi N b^2)]^{1/2} 
\end{equation}
In the above, $z$ defined by $z = x(2Nb^2)^{1/2}$  and $z_j = x_j(2Nb^2)^{1/2}$, $\nu  = —s{\tau_{R}}/2$ and $\Gamma(\nu)$ is the gamma function. Also, $z_{<}= min(z, z_0)$ and $z_{>}= max(z, z_0)$. $D_{\nu}$ represent parabolic cylinder functions. To get an understanding of the behavior of $k_I$ and $k_L$, we assume the initial distribution $P^0_e(x_0)$ is represented by $\delta(x-x_0)$. Then, we get 
\begin{equation}
{k_I}^{-1}= {k_s}^{-1}\left(1 - \frac{\frac{k_0}{\tau_{R}}F(z_s,k_s|z_0}{k_s+ {\frac{k_0}{\tau_{R}}}F(z_s,k_s|z_s)} \right)
\end{equation}
Again
\begin{equation}
k_L= k_r - [ values \; of \; s \; for \; which \;\; s+ {\frac{k_0}{\tau_{R}}} F(z_s,s|z_s)=0]
\end{equation}
We should mention that $k_I$ is dependent on the initial position $x_0$ and $k_s$ whereas $k_L$ is independent of the initial position.
In the following , we assume $k_s\rightarrow$ 0, in this limit we arrive at conclusions, which we expect to be valid even when $k_s$ is finite. Using the properties of $D_v{(z)}$, we find that when $k_s\rightarrow 0, F{(z_s,k_s|z_0)}$ and $F{(z_s,k_s|z_s)}\rightarrow exp(-z_s^2/2){[1/(4\pi Nb^2)]}^\frac{1}{2}$so that
\begin{equation}
\frac{k_0}{\tau_r} F{(z_s,k_s|z_0)}/[k_s+ \frac{k_0}{\tau_r}F{(z_s,k_s|z_s)}]\rightarrow 1.
\end{equation}
\\Hence 
\begin{equation}
k_I^{-1}=-{[\frac{\partial}{\partial k_s}\left[\frac{\frac{k_0}{\tau_R} F(z_s,k_s|z_0)}{k_s + \frac{k_0}{\tau_R} F(z_s,k_s|z_s)}\right]}_{k_s \rightarrow 0}
\end{equation} 
If we take $z_0 < z_s $, so that the particle is initially placed to the left of sink. Then \\
\begin{equation}
k_I^{-1}= \frac{e^{{z_s}^2/2} \tau_R}{k_0}{[1/{(4\pi N b^2)}]}^{1/2}+ \left[\frac{\partial}{\partial k_s}\left[\frac{e^{[(z_0^2-z_s^2)/4]}D_v{(-z_0)}}{D_v{(-z_s)}}\right]\right]_{v=0}
\end{equation} 
After simplification
\begin{equation}
k_I^{-1}= \frac{e^{{z_s}^2/2} \tau_R}{k_0}{[1/{(4\pi N b^2)}]}^{1/2}+ \left(\int_{z_0}^{z_s} dz e^{(z^2/2)}\left[1+erf(z/\sqrt{2}\right]\right)(\pi/2)({\tau_R/2})
\end {equation}
In the limit $\frac{k_0}{\tau_R}\rightarrow \infty$ the particle would be absorbed if it reaches $x_s$.In this limit, the first term on the right hand side would disappear.
Now we introduce the dimensionless rate constants $\bar k_I$ and $\bar k_L$ by $\bar k_I = ({\tau_R})k_I/2$ and  $\bar k_L = ({\tau_R})k_L/2$ and another dimensionless rate $\bar{k_0} = {\frac{k_0}{[1/2b\sqrt{2N}]}}$.

\begin{equation}
k_I^{-1}= \frac {e^{{z_s}^2/2}}{\bar {k_0}+(2\pi)^{1/2}} \left[\int_{z_0}^{z_s} dz e^{(z^2/2)}dz \left(1+erf(z/\sqrt{2}\right)\right]{(\pi/2)^{1/2}}
\end {equation}
When $ {\frac {k_0}{\tau_R}}$ is in small limit, $\bar{\frac {k_0}{\tau_R}}\ll 1$, the first term on the right hand side is significant. Thus $k_I$ is then independent of ${\tau_R}$. The short term rate exhibits Arrhenius type activation. Now if $ {\frac {k_0}{\tau_R}}$ is large then  $\bar{\frac {k_0}{\tau_R}\gg 1}$  is large, then $k_I$ shows inverse dependence on relaxation time ${\tau_R}$. This short term rate would increase with increase in length of polymer $N$ as well as bond length. Thus for rigid polymers, where bond length is small the short term rate constant is small and for flexible polymers the short term rate constant  would increase with square of the bond length. Thus maximum looping is achieved for a polymer of larger length and shorter bond length. 

\noindent The long-term rate constant $k_L$ is determined by the value of $s$, which satisfy $s+ \frac{k_0}{\tau_R}F(z_s,s|z_s)=0$. This equation can be written as an equation for $\nu (= -s{\tau_R} /2)$
\begin{equation}
\nu = D_\nu(-z_c)D_\nu(z_c)\Gamma(1-\nu)\frac{k_0}{4 b \sqrt{\pi N}} 
\end{equation}
For integer values of $\nu$, $D_\nu(z)=2^{-\nu/2}e^{-z^2/4}H_{\nu}(z/\sqrt{2})$, $H_{\nu}$ are Hermite polynomials. $\Gamma(1-\nu)$ has poles at $\nu = 1,2, . . . .$. Our interest is in $\nu \in [0, 1]$, as $k_L = \frac{2}{\tau_R} \nu$ for $k_s =0$. If $ \frac{k_0}{(4 b \sqrt{\pi N})}\ll 1$, or $z_c \gg 1$ then $\nu \ll 1$ and one can arrive
\begin{equation}
\nu = D_0(-z_s)D_0(z_s)\frac{k_0}{(4 b \sqrt{\pi N)}} 
\end{equation}
and hence 
\begin{equation}
k_L = \frac{e^{{-z_s}^2/2} \tau_R}{k_0}{[1/{(4\pi N b^2)}]}^{1/2}
\end{equation}
In this limit, the rate constant $k_L$ exhibits Arrhenius type activation. Thus the long term rate shows direct dependence on the relaxation time. It shows weaker dependence on the chain length as well as bond length of the polymer.
Relaxation time dependence of rate constants
\begin{table}[ht]
\caption{Relaxation time dependence of rate constants} 

\centering 
\begin{tabular}{c c c }
\hline\hline 
 & $k_{L}$  & $k_{I}$ \\ [0.5ex] 
\hline 
$\bar{\frac {k_0}{\tau_R}}\ll 1$  & 	Independent & Dependent \\ 
$\bar{\frac {k_0}{\tau_R}\gg 1}$  &     Fractional & Fractional \\
\hline 
\end{tabular}
\label{table:nonlin} 
\end{table}

\section{Conclusions:}
The model describes physical properties of polymer. To recapitulate, the relaxation time dynamics of end-to-end looping of polymer chains is sufficiently described by our simple polymer model. In the small limit
The short term rate constant shows an inverse dependence on the relaxation time but a direct dependence on the square root of chain length of the polymer $(N)$.Our calculation shows that for short term rate constant, relaxation time makes loop formation between two ends of a polymer chain slower but has a stronger dependence on the chain length as well as bond length of the polymer. Whereas for long term rate constant, relaxation time  We show that for shorter polymer strands looping is slower as compared to polymer having longer chain length. The bond length $(b)$ strongly affects the probability of loop formation. For long constructs of polymer, the end-to-end distance is found to exhibit inverse power law relationship to bond length. Interestingly, we show that these modifications alter... To gain insight,

\noindent In the present work  

\section{Acknowledgments:}

\end {document}